\documentclass[11pt,a4paper]{article}

\usepackage[numbers]{natbib}
\usepackage[T1]{fontenc}
\usepackage[utf8]{inputenc}
\usepackage{amsmath,amssymb}
\usepackage{array}
\usepackage{graphicx}
\usepackage{booktabs}
\usepackage{makecell}
\usepackage[table]{xcolor}
\usepackage{textcomp}
\usepackage{tikz}
\usepackage{xurl}
\usepackage[margin=1in]{geometry}
\IfFileExists{microtype.sty}{\usepackage{microtype}}{}
\usepackage[hidelinks]{hyperref}
\usetikzlibrary{arrows.meta,positioning,shapes.geometric}

\makeatletter
\g@addto@macro\UrlBreaks{\do\_\do\-\do\.\/}
\makeatother
\DeclareRobustCommand{\tool}[1]{\texttt{\nolinkurl{#1}}}
\DeclareRobustCommand{\status}[1]{\texttt{\nolinkurl{#1}}}
\newcolumntype{P}[1]{>{\raggedright\arraybackslash}p{#1}}
\newcolumntype{Y}[1]{>{\centering\arraybackslash}p{#1}}

\newcommand{\rev}[1]{#1}

\title{Closed-Loop Evaluation of LLM Agents for Embedded Software Development}

\author{%
\parbox{0.94\textwidth}{\centering
Jorge García-Carrasco\textsuperscript{1,*}\quad
Sergio García-Carrasco\textsuperscript{2}\quad
Alejandro Maté\textsuperscript{1}\quad
Juan Trujillo\textsuperscript{1}\\[1.2ex]
\textsuperscript{1}Lucentia Research, Department of Software and Computing Systems, University of Alicante,
Ctra. de San Vicente del Raspeig s/n, 03690 Sant Vicent del Raspeig, Spain\\
\textsuperscript{2}Universitary Institute of Materials Technology (IUTM), Universitat Politècnica de València,
Plaça Ferrándiz i Carbonell, 03801 Alcoi, Spain\\[0.8ex]
\textsuperscript{*}Corresponding author: \texttt{jorge.g@ua.es}\\[0.4ex]
ORCID: 0000-0003-3174-083X (J.~García-Carrasco), 0009-0001-9854-0309 (S.~García-Carrasco),
0000-0001-7770-3693 (A.~Maté), 0000-0003-0139-6724 (J.~Trujillo)\\[1.2ex]
{\small Published in \emph{Journal of Systems Architecture} 179 (2026) 103937,
\url{https://doi.org/10.1016/j.sysarc.2026.103937}. \copyright~2026 The Authors.
Published by Elsevier B.V. under the CC BY 4.0 license
(\url{https://creativecommons.org/licenses/by/4.0/}).}
}}

\date{}

\begin{document}

\maketitle

\begin{abstract}
\rev{Large language models (LLMs) are increasingly deployed as coding agents that edit files, run builds and tests, inspect execution results, and repair software iteratively. Embedded firmware is a demanding target because correctness depends on closed-loop behavior under sensing, timing, and safety constraints, not only on static source quality. Yet embedded-agent evaluation remains limited and often emphasizes one-shot synthesis or offline correctness.}

\rev{We present a benchmark for closed-loop evaluation of embedded coding agents. Each task provides a plain-text engineering description, constrained workspace, and visible build-and-runtime surface. The agent must translate requirements into implementation and self-verification steps, then iterate until the required device behavior is achieved. The suite contains five embedded-control tasks and four feedback scenarios: one-shot generation, realistic self-verification, CI-style red/green feedback, and oracle-style detailed feedback.}

\rev{The implementation targets simulated ESP32 firmware for reproducibility. We evaluate seven GPT-family and Qwen-family configurations across five tasks and four scenarios, with three repetitions per condition for 420 runs. \tool{gpt-5.4} has the highest pass rate among evaluated configurations but does not saturate the benchmark; \tool{qwen3.5-27B} is the strongest observed local model; and smaller local models degrade sharply in pass rate and search efficiency. These results suggest that capable local embedded coding agents are emerging.}
\end{abstract}

\noindent\textbf{Keywords:} embedded software, large language models, language model agents, software engineering benchmarks

\medskip
\noindent\textbf{Highlights.}
\begin{itemize}\setlength{\itemsep}{2pt}\setlength{\parskip}{0pt}
\item We introduce a closed-loop benchmark for embedded firmware coding agents, i.e., tool-using LLM systems that can autonomously execute bounded build-test-debug loops.
\item The benchmark targets why agents matter for embedded work: tasks begin from plain-text engineering handoffs and require visible build, test, runtime, repair, and stopping decisions.
\item The evaluation spans five tasks, four feedback regimes, seven models, and three repetitions for a total of 420 runs.
\item The model set combines two hosted frontier systems with five local Qwen variants.
\item \rev{\tool{gpt-5.4} has the highest observed pass rate among the evaluated configurations, while \tool{qwen3.5-27B} is the strongest observed local model and points to viable on-premises embedded-agent workflows.}
\end{itemize}

\section{Introduction}
Large language models (LLMs), trained on very large corpora of natural language and source code, now show strong capabilities across a wide variety of tasks \cite{kaplan2020scaling}. In software settings, they can interpret requirements, summarize repository state, generate candidate implementations, explain failures, and propose repairs \cite{roziere2023code}. By themselves, however, they usually act as single-turn text generators. Agentic coding systems extend them with a persistent workspace and explicit tool use \cite{react}, allowing the model to read and edit files, run builds and tests, inspect logs, and revise a partial solution over multiple steps rather than answering only once. In other words, a coding agent does not only suggest code; within a bounded tool environment, it can autonomously act on a repository, observe the consequences, and continue working until it decides to stop or submit.

\rev{This shift from one-shot generation to iterative, feedback-driven work is central to practical automation in software engineering. Recent benchmark work has shown that realistic measurement must account for whether models can process failure signals and improve a solution over repeated attempts \cite{swebench, swebench_verified}. Agent-oriented systems and evaluations further emphasize the importance of planning, tool use, execution, and iterative repair in realistic development environments \cite{sweagent, code_agent_eval}. The agent framing is therefore useful because many engineering tasks are bottlenecked not by producing an initial candidate, but by validating, debugging, and repairing that candidate until it satisfies the intended behavior. Embedded software is a particularly compelling target for this kind of evaluation: firmware correctness depends on behavior in closed loop with sensors, actuators, timing constraints, and safety interlocks, so success requires more than static source-level quality.}

Embedded software development is therefore qualitatively different from standard code-generation tasks. A controller may look reasonable in a code review and still fail once it interacts with asynchronous sensor streams, hysteresis, timeout handling, or sequence control. For the same reason, repository-level issue-resolution benchmarks do not fully cover the embedded setting either: they capture iterative software work, but not the hardware-facing closed loop in which firmware correctness depends on dynamic behavior.

Despite recent interest in applying modern coding agents to embedded and hardware-adjacent domains, the dominant emphasis is still on static synthesis, component-level checks, or offline correctness \cite{embedbench}. What remains less explored is the full agentic loop that makes modern coding agents useful in practice: starting from a plain-text specification of what must be implemented, formulating tests, implementing firmware, running it, learning from feedback, and iterating until the system behavior matches the requirements.

This motivates the central question of this paper:

\begin{center}
\colorbox{black!6}{%
\parbox{0.93\columnwidth}{%
Can an agent, starting from a plain-text specification of required device behavior, iteratively write tests and firmware, execute that firmware in a target environment, and converge to the required embedded behavior through a closed feedback loop?%
}}
\end{center}

We argue that this question deserves its own benchmark because the capability of interest is not one-shot firmware generation, but iterative embedded engineering. The benchmark we present is intended to evaluate whether an agent can start from a prose engineering brief, translate that brief into implementation and verification actions, and ultimately satisfy a behavioral contract in closed loop. The design is therefore prose-first, contract-aligned, and centered on the real development steps that embedded engineers actually perform: implementing controllers, writing or refining tests, running builds, probing runtime behavior, and iterating on failure signals.

In the current reference instantiation, we realize this methodology with an ESP32 firmware target running in QEMU \cite{systems_sim_eval} and deterministic plant simulators that model the controlled process. We simulate both the microcontroller and the plants/tasks because a benchmark should first be safe, repeatable, and easy to run at scale. This makes the evaluation reproducible and suitable for systematic comparison, while still exercising the core embedded loop of sensing, control, actuation, and verification. In this reference stack, the physical sensor/actuator boundary is intentionally virtualized through a UART protocol between firmware and simulator, which abstracts board-specific electrical integration while preserving the closed-loop control logic that the benchmark is meant to test. The ESP32/QEMU stack is therefore a concrete vehicle for the benchmark, not the boundary of the methodology itself.

\paragraph{Contributions.}
The paper makes the following contributions:
\begin{enumerate}
    \item A benchmark formulation for closed-loop evaluation of LLM agents on embedded firmware tasks, starting from a prose specification and ending in behavior-level validation.
    \item A benchmark design centered on contract-aligned task packets, visible self-verification affordances, a five-task difficulty ladder, and four evaluation scenarios with different degrees of feedback visibility.
    \item A reference ESP32/QEMU simulation instantiation with five deterministic plant models that enables safe and reproducible closed-loop evaluation before hardware deployment.
    \item A pooled 420-run evaluation across five tasks, four feedback scenarios, and seven models spanning hosted and local agents, showing that the benchmark is executable and model-separating, that \rev{\tool{gpt-5.4} has the highest observed pass rate without saturating the suite under repeated measurement}, and that \rev{\tool{qwen3.5-27B} is the strongest observed local operating point} while smaller local models degrade sharply in both accuracy and efficiency.
\end{enumerate}

The remainder of the paper is structured as follows. Section~\ref{sec:related} reviews related work. Section~\ref{sec:benchmark} presents the benchmark objective, contract, and design principles. Section~\ref{sec:evaluation} describes the evaluation setup and reports results across tasks, feedback scenarios, and models. Section~\ref{sec:discussion} discusses implications, limitations, and benchmark-design lessons. Section~\ref{sec:conclusion} concludes and outlines future work.

\section{Related Work}
\label{sec:related}

Prior work on software-engineering evaluation has established that realistic measurement requires more than one-pass code generation. Benchmarks such as SWE-Bench and its follow-ups showed the importance of testing whether models can interact with repositories, process failure signals, and revise solutions over multiple iterations \cite{swebench, swebench_verified}. In parallel, agent-oriented systems and evaluations such as SWE-agent \cite{sweagent}, Agentless \cite{code_agent_eval}, and related program-improvement frameworks have emphasized planning, tool use, execution, and iterative correction in realistic development environments \cite{autocode_agents}. Together, this line of work provides the general evaluation perspective we adopt: practical capability is not only the ability to synthesize code, but also the ability to use feedback productively in a development loop.

At the same time, synthesis-focused benchmarks such as HumanEval \cite{humaneval}, MBPP \cite{mbpp}, and RepoBench \cite{repobench} remain important because they isolate core functional code-generation ability. These datasets provide strong signal on local correctness and pattern-level coding skill. However, they do not fully represent embedded firmware work, \rev{where correctness often depends on stateful execution, asynchronous inputs, target constraints, low-level verification concerns \cite{manjunath2024rtos}, and behavior under disturbance.} In embedded settings, code that appears plausible in isolation may still fail once it is compiled, executed on a target, and exercised under temporal and safety constraints.

The literature on LLMs for embedded and hardware-adjacent programming has also grown rapidly. Recent benchmark work has explicitly targeted embedded-system development and microcontroller-oriented evaluation, helping establish firmware engineering as a serious target for modern models \cite{embedbench}. Complementary studies have benchmarked LLMs for embedded systems programming in microcontroller-driven IoT applications, further broadening the empirical base for the area \cite{embedagent}. Survey work has also reviewed transformer-based source-code generation more broadly, covering coding-model architectures, tuning strategies, datasets, and evaluation metrics \cite{ghaemi2024transformers_jsa}. Adjacent low-level work has examined code generation for hardware-description and synthesis-oriented settings, including Verilog-focused benchmarks that are relevant to the broader question of model capability in hardware-near engineering domains \cite{embedded_code_llm, hdl_bench}. \rev{Related systems work has also studied edge-device ML benchmarking \cite{baciu2024mlino} and on-device LLM deployment from a fixed-point simulation and hardware co-design perspective \cite{kim2025fixedpoint_jsa}, while agentic reasoning architectures have been explored for logic-code simulation tasks \cite{chen2026dcola_jsa}.} This work is valuable in establishing embedded software and related low-level domains as relevant targets for modern models. However, much of the existing evidence still centers on static synthesis, component-level checks, offline correctness, or narrow domain-specific reasoning tasks. Comparatively less attention has been given to the full agentic workflow in which a model starts from a prose engineering brief, writes or refines self-verification, runs firmware in a target environment, interprets staged feedback, and iterates until behavior-level requirements are satisfied.

\rev{Finally, simulation and emulation have long served as foundational evaluation substrates in robotics \cite{robotics_sim}, systems research \cite{systems_sim_eval}, and firmware analysis, including interrupt modeling from firmware logic \cite{wei2024iemu}, because they enable reproducibility, controlled experimentation, and cost-effective stress testing.} For embedded-agent research, simulation offers the practical advantage of repeated closed-loop execution without the fragility, safety risk, and operational overhead of immediate physical deployment. At the same time, simulation is not a replacement for real-device validation. In this paper, we use it as a deliberate methodological staging ground that supports rigorous comparison of agent behavior today, while informing future transitions to hardware-in-the-loop and real-machine studies.

Taken together, prior work motivates a benchmark that embraces the agentic evaluation workflow now common in academia and industry for realistic assessment of LLM agents in embedded development.

\section{Benchmark Design}
\label{sec:benchmark}

This benchmark evaluates whether an LLM agent can carry out a realistic embedded development loop from a natural-language engineering handoff. For each task, the agent receives a plain-text specification of required behavior and is expected to translate that specification into implementation and verification actions. In practice, this includes writing firmware, creating or refining tests, invoking builds and checks, interpreting failures, and iterating until behavior meets the stated requirements.

The benchmark is designed around simulated tasks so that experiments remain safe, repeatable, and comparable across models. Simulation allows many controlled runs with consistent task conditions while still exercising the core engineering workflow that matters for embedded automation. Detailed interfaces, execution mechanics, and evaluator behavior are introduced in the following subsections.

\subsection{Benchmark objective}

The benchmark is designed to evaluate whether an agent can complete an embedded firmware task from a realistic but controlled engineering handoff: a prose description of required device behavior, a constrained implementation surface, visible tool affordances, and a target environment that can be executed repeatedly. The emphasis is on agentic capability, not only on final-source quality.

More concretely, the benchmark is intended to answer three questions:

\begin{enumerate}
    \item Can the agent start from a simple plain-text specification and turn it into useful local verification, rather than depending on a benchmark-specific structured contract or hidden grader details?
    \item Can the agent iteratively write or refine firmware and tests, build target code, run local checks, execute the controller against a target environment, and use the resulting feedback to improve the implementation?
    \item Can the agent use that loop to reach a final correct submission that satisfies both nominal control requirements and safety-critical requirements such as stale-input handling, interlocks, timeout behavior, and sequence recovery?
\end{enumerate}

At a higher level, the benchmark is asking whether modern coding agents can do something that is both practically important and still underexplored in embedded software: begin from a prose engineering handoff, construct their own verification loop, and iterate all the way to behaviorally correct firmware rather than merely producing plausible source code. This is scientifically interesting because it moves the evaluation target from static synthesis to closed-loop engineering performance, and it is practically interesting because that is the level at which embedded automation would become genuinely useful in real development workflows.

In the current implementation, these questions are instantiated on an ESP32 firmware target running in QEMU \cite{systems_sim_eval} and coupled to deterministic plant models. This concrete stack is important because it enables reproducible, controlled comparisons across models. However, the primary scientific contribution is the benchmark methodology itself: a prose-first, closed-loop evaluation framework that can be transferred to other embedded targets and simulation environments.

\subsection{Benchmark workflow and contract}

\begin{figure*}[t]
\centering
\begin{tikzpicture}[
    font=\small,
    block/.style={draw=black, text=black, align=center, minimum height=8mm, minimum width=31mm},
    wideblock/.style={draw=black, text=black, align=center, minimum height=10mm, text width=45mm},
    innerblock/.style={draw=black, text=black, align=center, minimum height=8mm, text width=28mm},
    decision/.style={diamond, draw=black, text=black, aspect=2.1, align=center, inner sep=1pt, text width=17mm},
    io/.style={draw=black, text=black, trapezium, trapezium left angle=75, trapezium right angle=105, align=center, minimum height=8mm, text width=24mm},
    flow/.style={draw=black, -Latex, line width=0.8pt},
    guide/.style={draw=black, densely dotted, line width=0.8pt}
]
\draw[guide] (-6.0,3.45) -- (7.5,3.45);
\node[anchor=east, text=black] at (-6.05,3.45) {\small Input};
\draw[guide] (-6.0,-7.55) -- (7.5,-7.55);
\node[anchor=east, text=black] at (-6.05,-7.55) {\small Output};

\node[io] (brief) at (-2.05,4.45) {Task\\description};
\node[block] (agent) at (-2.05,1.95) {LLM Agent};
\node[wideblock] (local) at (-2.05,0.05) {\textbf{Local self-verification loop}: edit code, build firmware,\\run tests, boot QEMU, inspect probes};
\node[decision] (submit) at (-2.05,-2.15) {Submit?};
\node[block] (hidden) at (-2.05,-4.35) {Hidden evaluator};
\node[decision] (accept) at (-2.05,-6.15) {Accepted?};
\node[io] (pass) at (-2.05,-8.45) {Accepted\\firmware};
\node[io] (fail) at (3.05,-8.45) {Rejected\\submission};

\draw[black] (3.15,2.8) rectangle (6.85,-0.95);
\node[font=\small\bfseries, text=black] at (5.0,3.05) {Visible Workspace};
\node[innerblock] (edit) at (5.0,1.95) {Editable\\firmware + tests};
\node[innerblock] (tooling) at (5.0,0.05) {Build, QEMU,\\logs, traces};

\draw[flow] (brief) -- (agent);
\draw[flow] (agent) -- (local);
\draw[flow] (local) -- (submit);
\draw[flow] (submit) -- node[right, font=\scriptsize] {yes} (hidden);
\draw[flow] (hidden) -- (accept);
\draw[flow] (accept) -- node[right, font=\scriptsize] {yes} (pass);
\draw[flow] (accept.east) -| node[pos=0.22, above, font=\scriptsize] {no} (fail.north);
\draw[flow] (submit.west) -| ++(-1.4,0) |- node[pos=0.22, left, font=\scriptsize] {no} (agent.west);

\draw[flow] (agent.east) -- node[above, font=\scriptsize] {tool actions} (edit.west);
\draw[flow] (edit) -- (tooling);
\draw[flow] (tooling.west) -- node[pos=0.60, above, font=\scriptsize, fill=white, inner sep=1pt] {artifacts and logs} (local.east);
\end{tikzpicture}
\caption{Main agentic workflow evaluated by the benchmark. The agent iterates locally on visible evidence, explicitly decides whether to submit, and only then triggers a hidden behavioral evaluation.}
\label{fig:agentic-workflow}
\end{figure*}
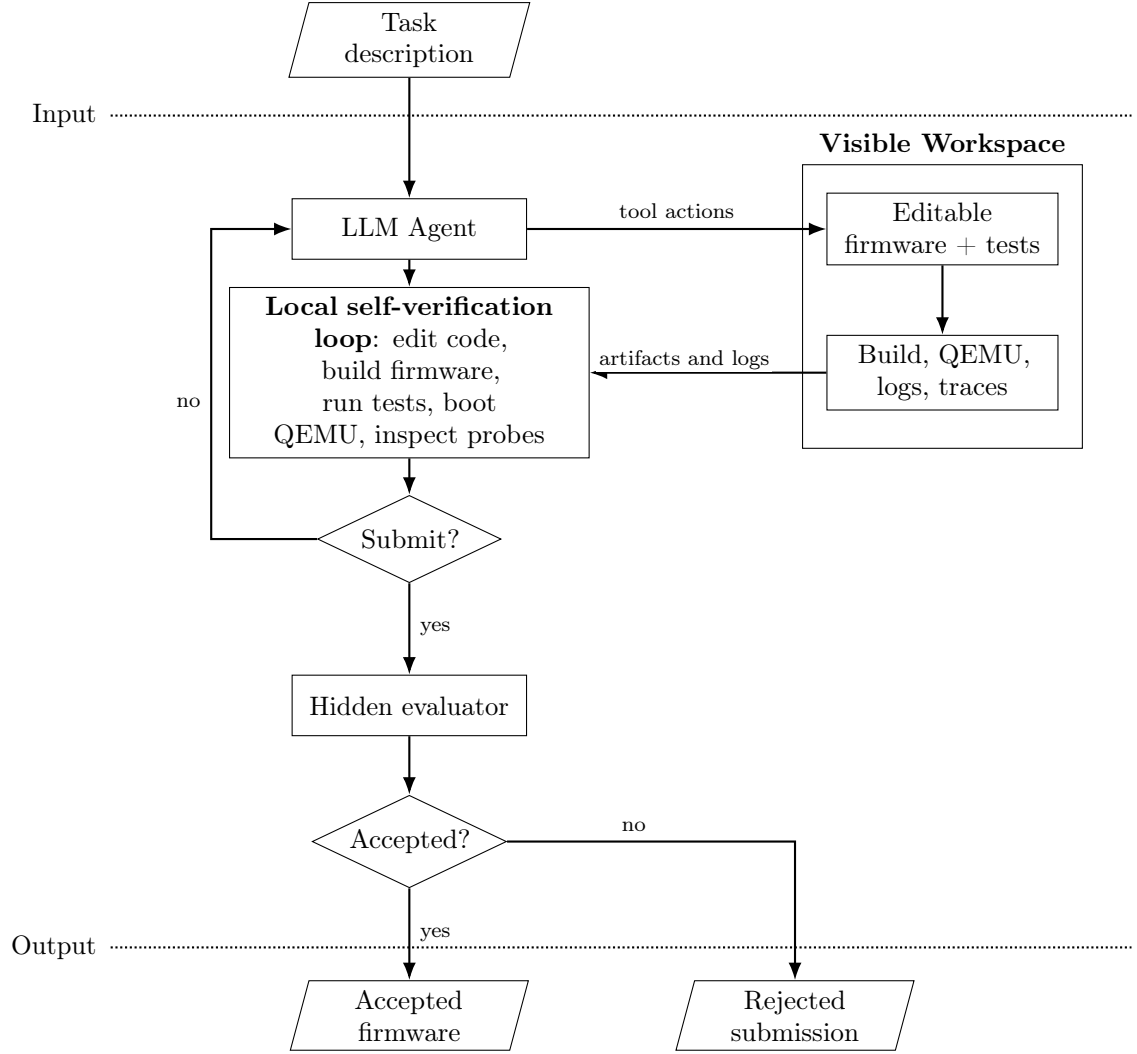

The primary workflow evaluated by this benchmark is the \tool{realistic_self_verify} setting, in which the agent receives a prose engineering brief, operates inside a constrained workspace, and uses visible local tooling to verify its work before deciding whether to submit to the hidden evaluator. In this primary condition, the hidden evaluator is not part of the local repair loop: the agent iterates on the basis of visible evidence, not hidden failure messages. Figure~\ref{fig:agentic-workflow} summarizes this workflow. The one-shot and richer-feedback settings described later in Section \ref{sec:evaluation_scenarios} are useful comparison conditions that will also be explored, but this self-verification-centered workflow is the main target of the benchmark.

Therefore, the workflow induces a simple contract. For each task:

\begin{enumerate}
    \item The agent receives a plain-text specification file describing the required behavior.
    \item The agent can edit only the intended implementation surface.
    \item The agent can repeatedly use visible build, test, and runtime tooling to decide whether to keep iterating locally or to submit.
    \item The hidden evaluator is consulted only upon submission; in the primary workflow it does not provide detailed failure feedback to drive the local loop.
    \item The task is solved only when a submitted implementation satisfies the full behavioral contract.
\end{enumerate}

This contract is deliberate. The goal is to measure whether an agent can start from a natural engineering handoff and iterate until the device behavior meets that handoff. The benchmark is therefore prose-first by default: the internal structured task definition remains hidden, and familiarity with benchmark-specific schemas is not supposed to be the main source of success.

\subsection{Design principles and construction process}

The workflow above describes what the agent experiences during evaluation. We next explain how the benchmark itself is built so that difficulty comes from the embedded control problem rather than from accidental benchmark artifacts. The following design principles guided both task design and the construction of the evaluator, simulators, and public task packets.

The benchmark follows five design principles.

\begin{enumerate}
    \item \textbf{Closed-loop first:} The benchmark should evaluate iterative development, not just final code emission.
    \item \textbf{Contract alignment:} Hidden checks must trace back to visible requirements in the prose handoff. If a hidden assertion matters, it should be stated in the public task packet; otherwise it should not be graded.
    \item \textbf{Task-specific failures:} If an agent fails, the failure should be attributable to the control problem, such as a wrong state transition, stale-sensor policy, or timeout rule, rather than to benchmark setup errors such as incorrect build instructions in the task handoff.
    \item \textbf{Visible self-verification affordances:} Realistic self-verification requires visible tooling, test locations, and runtime hooks; merely telling the agent to verify itself is not enough.
    \item \textbf{Stack portability:} The benchmark design should remain conceptually independent of the current microcontroller and simulator choices.
\end{enumerate}

These principles are aimed at benchmark validity. A failure should be interpretable as a failure on the embedded-development problem, not on hidden benchmark trivia. Likewise, a success should reflect genuine convergence in a closed loop rather than accidental matching against an overexposed grading contract.

We operationalize these principles through a fixed construction pipeline for every task in the suite so that task difficulty comes from the embedded control problem, not from accidental benchmark artifacts. The process begins with task design: first, an engineering-level control plan is written so that it specifies intended behavior, safety constraints, off-nominal cases, and acceptance criteria. This plan is the source for both the public handoff and the hidden evaluator, which keeps the benchmark contract aligned across visible and private components.

The next stage is simulation and execution-harness construction. For each task, we implement a deterministic process simulator and connect it to the firmware target so that the controller is evaluated in a true closed loop. As previously mentioned, in the current reference stack, the firmware runs on an ESP32 target in QEMU, and the process simulator exchanges sensor and actuator signals with that target through UART. We use simulation first because it is safer, repeatable, and practical for large comparative sweeps, while still preserving the core embedded workflow of sensing, control, actuation, and runtime feedback.

After the simulator is stable, we write hidden ground-truth tests for grading. These tests are derived from the same requirements and cover nominal trajectories, boundary conditions, and safety-critical behaviors such as stale-input handling, timeout policies, and interlocks. The hidden suite defines final pass/fail status and remains private so that models cannot overfit to evaluator internals. Hidden-evaluator feedback is returned only in the two feedback-rich evaluation scenarios to assess whether this extra information helps the agents converge better and faster.

In parallel, we prepare the agent-facing task packet. The agent receives a plain-text natural-language handoff, not a machine-readable contract and not the hidden tests. The workspace exposes the required tooling for build, local verification, and runtime probing, but editing is constrained to the intended implementation surface. This constraint is enforced by the harness so that comparisons remain fair and focused on the designated firmware-development scope.

During evaluation, the agent must turn the prose brief into executable verification and implementation steps: writing or refining tests, modifying firmware, running builds, interpreting failures, and iterating in simulation. The harness records staged outcomes such as logic-test failure, build failure, runtime failure, or behavioral failure, and final success is determined only by the hidden ground-truth evaluator. This design simultaneously measures agentic development behavior and end-to-end behavioral correctness.

\subsection{Task suite}
\label{sec:tasks}

The current task suite contains five tasks arranged as a deliberate difficulty ladder. Each task starts from a plain-text specification of desired behavior and requires the agent to converge to a correct controller through iterative validation. Difficulty is intended to come from the control problem itself: additional sensing channels, stronger coupling between decisions, explicit safe-state behavior, and eventually a full sequence-control problem. The defined tasks are presented below:

\begin{table*}[t]
\caption{Compact view of the five-task difficulty progression.}
\label{tab:task-ladder}
\small
\begingroup
\renewcommand{\arraystretch}{1.0}
\begin{tabular*}{\textwidth}{@{}cP{0.31\textwidth}P{0.56\textwidth}@{}}
\toprule
Tier & Task & Primary added complexity \\
\midrule
1 & \tool{tank_fill_drain} & Baseline threshold control with timeout-safe behavior. \\
\addlinespace[4.5pt]
2 & \tool{thermal_chamber_hysteresis} & Adds plant dynamics and anticipatory control near the upper band. \\
\addlinespace[4.5pt]
3 & \tool{pressure_vessel_interlock} & Adds asynchronous input freshness and explicit interlock safety constraints. \\
\addlinespace[4.5pt]
4 & \tool{mixing_tank_fill_heat} & Adds multi-sensor coupling and phase-dependent control decisions. \\
\addlinespace[4.5pt]
5 & \tool{filter_tank_sequence} & Adds timer-driven sequence control, evidence accumulation, and recovery behavior. \\
\bottomrule
\end{tabular*}
\endgroup
\end{table*}

\begin{enumerate}
    \item \textbf{\tool{tank_fill_drain}:} A baseline single-sensor task with one level stream and one pump actuator. The task brief asks for simple threshold control, but not stateless threshold control: the controller must turn the pump on below 30, turn it off above 80, and preserve the previous safe output inside the 30--80 band rather than forcing a new command in the middle region. It must also distinguish normal short idle gaps from true timeout behavior and fail safe to pump-off on malformed input or stale data beyond 1000 ms. This task therefore establishes the basic pattern of prose specification, exact UART-visible actuation, timeout handling, and runtime self-verification.

    \item \textbf{\tool{thermal_chamber_hysteresis}:} A single-loop temperature-control task with a laggy plant and binary heater actuation. The controller must heat below 47, turn off above 52, preserve the previous safe state inside the nominal band, and perform anticipatory shutoff when temperature is rising in the 50--52 region. The prose brief explicitly treats that anticipatory heater-off as a valid holding behavior rather than a fault, so the agent must reason about trend, not only instantaneous thresholding. The task also includes malformed-input and stale-sensor fail-safe behavior plus a closed-loop hard ceiling at 56\,\textdegree C, which means a controller can satisfy the visible threshold logic and still fail if it ignores plant lag.

    \item \textbf{\tool{pressure_vessel_interlock}:} A dual-sensor interlock task combining pressure regulation with door safety logic. Pressure freshness and door freshness are independent, and the controller must remain in the safe vent-open state until both signals have been observed validly. Once initialized, it must pressurize below 40 when sealed, relieve above 60, preserve the previous safe output inside the band, and immediately yield to door-open dominance. The task also requires that the controller never command compressor-on and vent-open simultaneously, making it the first task in the ladder where explicit actuator mutual exclusion and asynchronous input validity both matter.

    \item \textbf{\tool{mixing_tank_fill_heat}:} A coupled level-and-temperature task with independent sensor freshness and a low-fill heater guard. The controller must open the inlet below 60, close it above 75, keep the heater off whenever level falls below 55, and only heat in the safe fill window when temperature is still below the working band. Near the upper temperature edge, it must again reason about temperature trend and switch the heater off early when temperature is rising. The task therefore combines multi-sensor validity, coupled decisions across two actuators, and a closed-loop hard ceiling at 51\,\textdegree C.

    \item \textbf{\tool{filter_tank_sequence}:} A sequence-control task with two sensors, two actuators, and explicit stage transitions for filtering, settling, draining, and completion. The controller must count a clear streak using valid turbidity inputs only, enter settling after three clear samples while level remains high enough, wait at least 400 ms before opening the drain, and then recover correctly from disturbances that arrive during settling or draining by restarting filtering and resetting progress. It must also detect completion when the remaining level drops below the minimum threshold and return to the all-off state. This is the strongest task in the current ladder because correctness depends on state progression, timing, evidence accumulation, and disturbance recovery rather than on a single local threshold rule.
\end{enumerate}
Taken together, these five tasks are intended to form a deliberate ladder of increasing embedded-control difficulty, as summarized in Table~\ref{tab:task-ladder}. The progression starts with a simple single-sensor threshold controller, then adds plant lag and anticipatory control, asynchronous multi-sensor interlocks, coupled multi-actuator decisions, and finally timed sequence control with explicit disturbance recovery. This structure is useful because it helps separate models not only by raw coding ability, but also by their ability to maintain controller state, reason over asynchronous observations, and build stronger self-verification loops as the control problem becomes richer.

\subsection{Evaluation scenarios}
\label{sec:evaluation_scenarios}

Although the primary research target is the setting in which the agent must build and trust its own verification loop, we report three additional scenarios to contextualize performance. Concretely, we keep one simpler one-shot baseline and two guidance-oriented settings so that results can be interpreted across a controlled range of feedback assumptions. The four evaluation scenarios are presented below, ordered by increasing level of feedback:

\begin{enumerate}
    \item \textbf{\tool{oneshot_blind}:} The agent is restricted to a single submission, no hidden feedback is returned, and the public self-verification helpers are intentionally omitted. This removes iterative search almost entirely and provides a lower-bound baseline for direct specification-to-code performance. In other words, it asks what happens when the model is forced to rely almost entirely on its first-pass interpretation of the task packet.
    
    \item \textbf{\tool{realistic_self_verify}:} This is the primary benchmark condition. The agent may iterate freely, but hidden grader details are not returned after submission. The agent must therefore rely on the visible build, test, and runtime surface exposed in the workspace and on whatever self-authored checks it creates under that surface. This is the closest match to the scientific question of the paper: can the agent read a prose engineering brief, construct its own verification loop, and decide for itself when the solution is ready to submit?

    \item \textbf{\tool{ci_red_green}:} The agent may iterate and resubmit, but after each submission it sees only a compact hidden pass/fail outcome. This approximates a CI-style search signal: the model learns whether the current attempt is acceptable, but not which hidden assertion failed or what concrete repair is needed. It is therefore useful for separating models that benefit from minimal external search guidance from models that can already verify themselves locally.

    \item \textbf{\tool{oracle_full}:} The agent may iterate and receives detailed hidden host-test or integration feedback after submission. This is intentionally more guidance-heavy than the intended autonomous-use case, but it is a useful diagnostic upper bound on repair under rich supervision. If a model succeeds only here, then the benchmark is showing that it can repair when pointed toward the error, but not necessarily that it can construct a trustworthy self-verification loop on its own.
\end{enumerate}

These four scenarios should therefore not be read as a single monotone leaderboard. They are complementary lenses over the same tasks. \tool{realistic_self_verify} measures the core target capability, \tool{oneshot_blind} isolates first-pass implementation skill, and \tool{ci_red_green} plus \tool{oracle_full} show how strongly each model depends on progressively richer hidden guidance. Together, they make it possible to attribute gains to different mechanisms: specification understanding, self-verification quality, and dependence on external feedback.

\rev{The scenario axis also provides a targeted ablation of the two design principles that are behavioral rather than benchmark-validity constraints: closed-loop iteration and visible self-verification. In particular, the contrast between \tool{oneshot_blind} and \tool{realistic_self_verify} holds the task prose, hidden evaluator, tool harness, model set, and run budget fixed while changing whether the agent can use the visible build, test, and runtime surface for iterative self-verification before submission. By contrast, contract alignment, task-specific failure attribution, and stack portability are benchmark-hygiene constraints enforced by construction: violating them would introduce hidden unstated requirements, harness artifacts, or stack-specific assumptions rather than isolate useful agent behavior.}

\subsection{Implementation details}

All code and data used to replicate the results of this paper are publicly available at \url{https://github.com/jgcarrasco/closed_loop_evaluation_agents_embedded}. The present implementation instantiates the benchmark on an ESP32 firmware target executed in QEMU and coupled to deterministic plant models implemented in Python. ESP-IDF provides the firmware build system and target support \cite{esp_idf}, QEMU provides a reproducible execution substrate for the embedded target \cite{systems_sim_eval}, and the open-source \tool{pi} coding-agent harness handles model interaction and tool mediation \cite{pi_mono_coding_agent}. For the local open-weight experiments, Docker provides reproducible environment packaging \cite{docker}, and \tool{llama.cpp} serves the local models under a common inference stack \cite{llamacpp}. Each task has a task-specific runtime module under \tool{sim/tasks/} that defines the hidden scenarios, advances the plant state, emits sensor frames, consumes actuator commands from the firmware, and evaluates whether the resulting closed-loop behavior satisfied the scenario contract. This keeps the benchmark logic explicit and inspectable for humans while keeping the hidden evaluator itself out of the agent-visible workspace.

The common runtime harness launches the firmware in QEMU, exposes \tool{UART0} through a local TCP socket, and communicates with the firmware over newline-delimited ASCII frames. On the plant-to-firmware side, the runtime sends task-specific sensor frames such as \texttt{SENSE LEVEL 25}, \texttt{SENSE TEMP 47}, or \texttt{SENSE DOOR OPEN}, plus explicit fault frames such as \texttt{FAULT SENSOR\_TIMEOUT}. On the firmware-to-plant side, the harness expects exact task-specific actuator lines such as \texttt{ACT PUMP ON}, \texttt{ACT HEATER OFF}, or \texttt{ACT DRAIN OPEN}. This UART boundary is a simulation surrogate for real embedded I/O, where the same controller logic would more typically read GPIO/ADC/bus-connected sensors and drive actuators through GPIO, PWM, relays, or peripheral interfaces. The benchmark intentionally abstracts that board-specific electrical integration layer so that the evaluation stays focused on closed-loop control behavior, staged verification, and repair rather than on one particular hardware wiring stack. This exact-string contract is therefore part of the benchmark: hidden checks do not grade internal helper names or refactors, but they do grade the externally visible control behavior and UART-visible actuation surface.

\rev{We use four outcome labels when summarizing hidden-evaluation results. \texttt{PASS} means that the submitted implementation satisfies all hidden host and closed-loop checks. \texttt{HOST} denotes a hidden host-test failure: the submission reaches the hidden evaluator, but fails host-side build or unit tests before the QEMU closed-loop plant scenarios are accepted. \texttt{INT} denotes an integration failure: the submission passes the preceding host-test stage but fails during the QEMU-based closed-loop evaluation, for example because the firmware emits the wrong UART-visible actuator command, mishandles timing or stale inputs, violates a safety constraint, or fails a task-specific plant scenario. \texttt{NONE} means that the run ended without a final submission.}

The plant models are deterministic state machines rather than stochastic simulators. For the simplest \tool{tank_fill_drain} task, for example, the plant state contains only the current level, the pump state, and simulated time in milliseconds. Each plant step advances time by a fixed tick and updates the level according to a simple fill-minus-drain rule. More complex tasks add lag, coupled sensors, multi-actuator dynamics, or explicit sequence state, but they follow the same principle: identical inputs should produce identical traces so that runs remain reproducible across models and across repeated evaluations. During each hidden scenario, the runtime records a full transcript of UART exchanges plus structured trace samples and summary metrics, which makes failures diagnosable without revealing the hidden checks to the agent.

\subsubsection*{Illustrative example}

The easiest task, \tool{tank_fill_drain}, is a good concrete example of the benchmark contract. The public task brief tells the agent that the controller must turn the pump on when \tool{LEVEL < 30}, turn it off when \tool{LEVEL > 80}, preserve the previous safe output in the mid-band, and fail safe to pump-off on malformed input or timeout. It also explicitly instructs the agent to create its own visible tests and at least one firmware-level runtime probe before submission. The hidden host checks then validate whether those public requirements were actually met. Representative hidden checks for this task include preserving the current safe \tool{ON} state in the mid-band and forcing pump-off after the bounded timeout window.

This example illustrates what the agent must actually do in the benchmark. It is not enough to emit plausible controller code. The agent must edit the constrained firmware surface, write or refine local tests, build the firmware, and ideally exercise it end to end through the UART-visible runtime path before deciding to submit. That distinction matters because an agent can easily produce false greens with incomplete local tests. For example, it may validate the obvious low-threshold and high-threshold behavior while missing the mid-band latch or the exact timeout reaction. Hidden tests and hidden closed-loop scenarios are therefore necessary to distinguish genuine behavioral correctness from a locally convincing but incomplete self-verification story.

\section{Evaluation}
\label{sec:evaluation}

We evaluate whether current coding agents can turn prose task descriptions into behaviorally correct embedded firmware under repeated closed-loop interaction. The study spans seven models across five tasks and four feedback scenarios, repeated three times for a total of 420 runs. The model set includes two hosted frontier systems (\tool{gpt-5.4} and \tool{gpt-5.4-mini}) and five local open-weight Qwen variants (\tool{qwen3.5-27B}, \tool{qwen3.5-35B-A3B}, \tool{qwen3.5-9B}, \tool{qwen3.5-4B}, and \tool{qwen3.5-2B}).

All model evaluations reported in this paper are executed through the same \tool{pi} harness \cite{pi_mono_coding_agent}. This is deliberate: keeping the agent wrapper, tool interface, prompt surface, and run logging fixed reduces the risk of attributing scaffold differences to model differences. We chose this model set to cover a useful spread of deployment settings, model scales, and architectural forms, so that the evaluation captures both cross-family capability differences and within-family local scaling behavior.

\rev{The hosted GPT models are included as frontier-scale reference points rather than as an exhaustive hosted-model leaderboard. The primary purpose of the model set is to evaluate closed-loop embedded-agent behavior across the selected configurations, with particular attention to whether locally served open-weight models can solve the tasks under the same harness. We therefore scope model-ranking claims to the evaluated configurations. Additional hosted baselines, including Claude Opus, would be valuable future work but are outside the controlled model set studied here.}

We report the results in three complementary analyses:

\begin{enumerate}
    \item First, we treat \tool{realistic_self_verify} as the primary condition because it most closely matches the autonomous closed-loop workflow the benchmark is intended to measure.
    \item Second, we use the full four-scenario matrix to study how feedback visibility changes model behavior.
    \item Third, we focus on the local Qwen variants and study the trade-offs between model size, performance, and computational cost.
\end{enumerate}

Regarding the local Qwen models, they were all executed on a single NVIDIA RTX 4090 using \tool{llama.cpp}. The evaluated family includes \tool{qwen3.5-27B}, \tool{qwen3.5-35B-A3B}, \tool{qwen3.5-9B}, \tool{qwen3.5-4B}, and \tool{qwen3.5-2B}. We intentionally kept this study within a common GGUF 4-bit deployment regime, rather than mixing quantized and higher-precision checkpoints, even though a smaller model such as 9B could fit at higher precision on the same hardware. All local Qwen runs were executed with thinking disabled. This keeps numeric precision, serving format, and explicit test-time reasoning mode from becoming additional moving variables in the local scaling study so that the within-family comparison remains methodologically controlled. The local GGUF checkpoints were obtained from Unsloth's published Qwen3.5 releases and deployment guide \cite{unsloth_qwen35}, which provide a consistent public source for the family-wide local models. We did not evaluate models larger than the 35B version because of hardware constraints. We also omitted the 0.8B model because the 2B model finished with zero passes across all of its runs, recorded a first submission in only eight of them in the final curated bundle, and otherwise spent most of its budget in nonproductive search. On that basis, a 0.8B run was unlikely to add useful discriminative evidence beyond extending the same failure regime.

Across all runs we report hidden outcome, failure family, stage reached, total tokens, tool calls, hidden-evaluation count, build attempts, self-test runs, and runtime-probe behavior. Runs were allowed to iterate until success or a one-hour wall-clock limit; \tool{oneshot_blind} and \tool{realistic_self_verify} remained restricted to a single submission, whereas the other scenarios allowed repeated submissions within that same time budget. Because the plant and harness are deterministic, the three repetitions are intended to measure agent/model stochasticity under a fixed environment rather than randomized task variation. \rev{This repetition count is intended to expose major behavioral differences and recurring failure modes, not to resolve small differences between adjacent model configurations. In each per-model/per-scenario analysis, the effective binary sample size is 15 outcomes, from five tasks and three repetitions. A normal-approximation two-proportion power calculation at two-sided $\alpha=0.05$ and 80\% power gives minimum detectable absolute pass-rate differences of approximately 0.46, 0.49, 0.49, and 0.44 for baseline pass rates of 0.1, 0.2, 0.3, and 0.5, respectively; at a 0.7 baseline, even a perfect 1.0 comparator remains below 80\% power under this approximation. We therefore treat close model orderings as descriptive trends, while emphasizing larger effect-size patterns across feedback regimes, failure classes, and local-versus-hosted configurations.} Because the OpenAI models run on hosted backends whereas the Qwen models are served locally, wall-clock is not used as a fairness-critical metric in cross-backend comparisons. It is used only in the local-only subsection, where all models share the same hardware and serving stack.

\rev{Table~\ref{tab:inference-config} summarizes the inference and harness configuration used in the evaluation. For hosted GPT models, we used the GitHub Copilot subscription access path to reflect a realistic developer-facing workflow; in that setting, sampling parameters are provider-managed and temperature/top-p controls are not exposed to the harness. For local Qwen models, the \tool{llama.cpp} server was configured following the non-thinking Qwen3.5 guidance from Unsloth \cite{unsloth_qwen35}. Submission limits were scenario-specific: one submission for \tool{oneshot_blind} and \tool{realistic_self_verify}, and repeated submissions within the wall-clock budget for \tool{ci_red_green} and \tool{oracle_full}. At the benchmark-harness level, we imposed no separate tool-call, message-turn, or iteration-count cap beyond the one-hour wall-clock budget and the scenario-specific submission budget.}

\begin{table}[t]
\caption{\rev{Inference and harness configuration used for the evaluated models.}}
\label{tab:inference-config}
\small
\begingroup
\renewcommand{\arraystretch}{1.08}
\begin{tabular}{@{}P{0.32\columnwidth}P{0.30\columnwidth}P{0.30\columnwidth}@{}}
\toprule
Parameter & Hosted GPT models & Local Qwen3.5 models \\
\midrule
Runtime & GitHub Copilot subscription via \tool{pi} & \tool{llama.cpp} via \tool{pi} \\
Temperature & Provider-managed & 0.7 \\
Top-p & Provider-managed & 0.8 \\
Top-k & Provider-managed & 20 \\
Min-p & Provider-managed & 0.0 \\
Presence penalty & Provider-managed & 1.5 \\
Repeat penalty & Provider-managed & 1.0 \\
Reasoning/thinking & Provider-managed & Disabled \\
Run budget & One hour per run & One hour per run \\
Submission budget & Scenario-specific & Scenario-specific \\
\bottomrule
\end{tabular}
\endgroup
\end{table}

\subsection{Primary Results: Realistic Self-Verification}

The benchmark's main scientific target is the \tool{realistic_self_verify} condition, in which the agent must construct and trust its own local evidence without hidden failure messages. This is therefore the primary comparison for interpreting model capability.

\begin{table*}[t]
\caption{Pooled primary-scenario results for \tool{realistic_self_verify} over three repetitions. For all listed models, this corresponds to 15 runs. The pass column reports pooled counts, pass rate, and Wilson 95\% confidence interval. Cross-backend comparison emphasizes interaction cost rather than wall-clock.}
\label{tab:realistic-overall}
\footnotesize
\begingroup
\renewcommand{\arraystretch}{1.0}
\begin{tabular*}{\textwidth}{@{}P{0.22\textwidth}ccccc@{}}
\toprule
Model & Passes & \makecell[c]{Avg total\\tokens} & \makecell[c]{Avg tool\\calls} & \makecell[c]{Avg hidden\\evals} & \makecell[c]{Avg self-test\\runs} \\
\midrule
\tool{gpt-5.4} & \makecell[c]{14/15 (93.3\%)\\[-1pt]{\scriptsize[70.2, 98.8]}} & 0.42M & 35.5 & 1.0 & 2.6 \\
\addlinespace[4.5pt]
\tool{gpt-5.4-mini} & \makecell[c]{8/15 (53.3\%)\\[-1pt]{\scriptsize[30.1, 75.2]}} & 1.03M & 54.7 & 1.1 & 4.4 \\
\addlinespace[4.5pt]
\tool{qwen3.5-27B} & \makecell[c]{9/15 (60.0\%)\\[-1pt]{\scriptsize[35.7, 80.2]}} & 6.90M & 109.0 & 1.5 & 12.2 \\
\addlinespace[4.5pt]
\tool{qwen3.5-35B-A3B} & \makecell[c]{4/15 (26.7\%)\\[-1pt]{\scriptsize[10.9, 52.0]}} & 4.40M & 92.5 & 1.5 & 16.1 \\
\addlinespace[4.5pt]
\tool{qwen3.5-9B} & \makecell[c]{2/15 (13.3\%)\\[-1pt]{\scriptsize[3.7, 37.9]}} & 6.76M & 113.8 & 6.6 & 38.5 \\
\addlinespace[4.5pt]
\tool{qwen3.5-4B} & \makecell[c]{2/15 (13.3\%)\\[-1pt]{\scriptsize[3.7, 37.9]}} & 6.14M & 114.9 & 12.7 & 37.7 \\
\addlinespace[4.5pt]
\tool{qwen3.5-2B} & \makecell[c]{0/15 (0.0\%)\\[-1pt]{\scriptsize[0.0, 20.4]}} & 77.39M & 987.5 & 0.0 & 4.2 \\
\bottomrule
\end{tabular*}
\endgroup
\end{table*}

\rev{Table~\ref{tab:realistic-overall} shows the clearest descriptive ordering in the paper.} \tool{gpt-5.4} leads the primary condition with 14/15 passes (93.3\%). Among the non-frontier models, the local \tool{qwen3.5-27B} posts the highest pooled count at 9/15 (60.0\%), narrowly ahead of \tool{gpt-5.4-mini} at 8/15 (53.3\%). The remaining local models are substantially weaker in this primary condition: \tool{qwen3.5-35B-A3B} reaches 4/15 (26.7\%), \tool{qwen3.5-9B} and \tool{qwen3.5-4B} each reach 2/15 (13.3\%), and \tool{qwen3.5-2B} remains at 0/15 (0\%).

This makes two points immediately visible. First, the benchmark is not frontier-saturated: even the strongest model leaves one miss in the primary condition, so the current five-task suite still produces visible headroom at the top end. Second, local agents are not merely toy baselines: the 27B local Qwen model posts a slightly higher pooled pass count than the smaller hosted frontier model in the benchmark's primary setting, although the overlapping Wilson intervals show that this gap should not be over-read as a stable separation. That competitiveness comes at a substantial search cost, however. Relative to \tool{gpt-5.4-mini}, \tool{qwen3.5-27B} uses roughly 6.7$\times$ as many tokens and about 2$\times$ as many tool calls per primary-condition run to obtain its additional pass. \rev{Token totals are the provider/server usage totals recorded by the \tool{pi} harness, including reported prompt/input, completion/output, and cache tokens where available. Because tokenization and reasoning-token accounting are backend-dependent, we interpret token ratios as reported interaction volume rather than provider-independent compute; tool-call ratios are the cleaner cross-backend behavioral signal.}

\begin{figure*}[t]
\centering
\includegraphics[width=\textwidth]{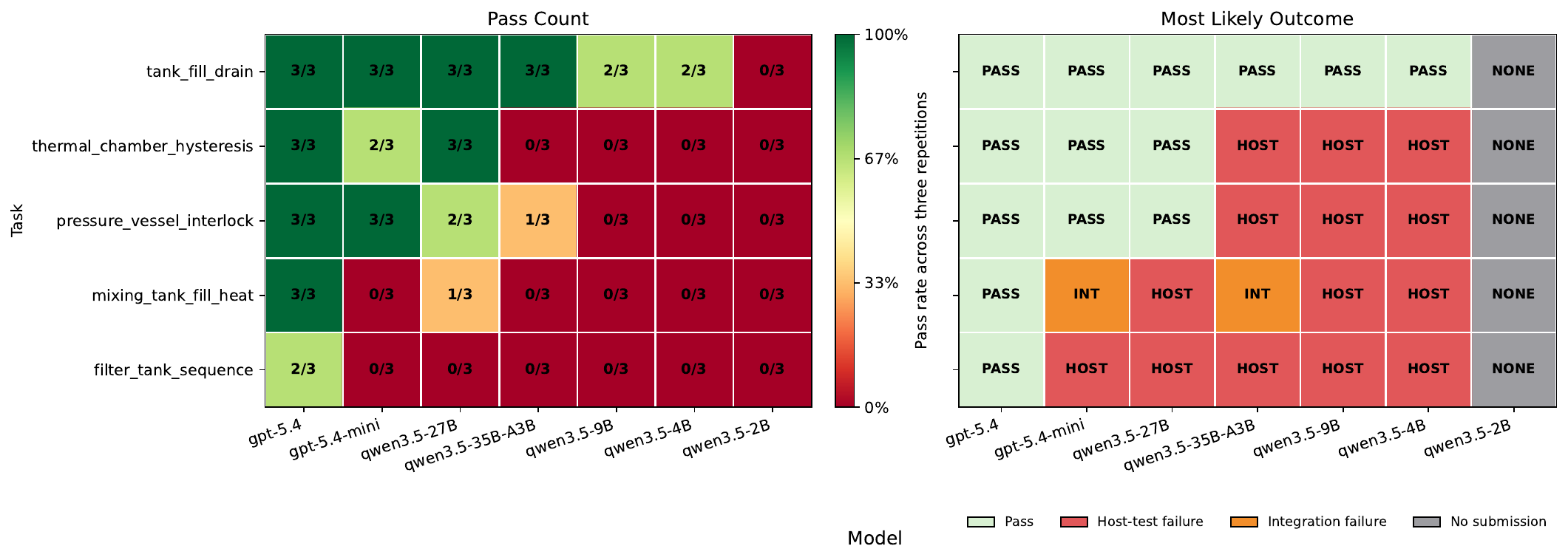}
\caption{Two-panel view of the primary \tool{realistic_self_verify} condition over three repetitions. The left panel shows pass counts out of three repetitions for each model-task pair. The right panel shows the most likely outcome across those repetitions: \texttt{PASS} if pass is the modal outcome, \texttt{HOST} for host-test failure, \texttt{INT} for integration failure, and \texttt{NONE} for no submission.}
\label{fig:realistic-by-task}
\end{figure*}

The left-hand panel in Figure~\ref{fig:realistic-by-task} reports pass counts out of three repetitions for each model-task pair, so it makes the replicated task-by-task separation explicit rather than relying only on the aggregate totals in the previous Table~\ref{tab:realistic-overall}.

\rev{Its main result is that the benchmark's observed ordering is carried by a small number of sharply discriminative tasks.} \tool{filter_tank_sequence} is the strongest separator: only \tool{gpt-5.4} passes it in the primary condition, and even then only in two of the three repetitions. At the other extreme, \tool{tank_fill_drain} is the anchor task: the two hosted models plus \tool{qwen3.5-27B} and \tool{qwen3.5-35B-A3B} solve it in every repetition, and even the weaker local models still manage it at least twice. The middle three tasks create the real spread. \tool{qwen3.5-27B} is the only local model that passes \tool{mixing_tank_fill_heat}, and it is also the strongest local model on \tool{thermal_chamber_hysteresis} and \tool{pressure_vessel_interlock}.

The right-hand panel reports the modal outcome for each cell across the three repetitions, using the outcome labels defined in Section~3.6 to distinguish stable passes from the most common failure regime: host-test failure, integration failure, or no submission.

That outcome view shows that the failure regimes are not uniform across models. \tool{gpt-5.4} is still overwhelmingly pass-dominated, with only one integration-style miss on \tool{filter_tank_sequence}. \tool{gpt-5.4-mini} passes \tool{tank_fill_drain}, \tool{thermal_chamber_hysteresis}, and \tool{pressure_vessel_interlock} in most repetitions, but \tool{mixing_tank_fill_heat} becomes a consistent integration failure and \tool{filter_tank_sequence} a consistent host-test failure. For \tool{qwen3.5-27B}, the losses are concentrated on the two hardest tasks: \tool{tank_fill_drain} and \tool{thermal_chamber_hysteresis} remain stable passes, \tool{pressure_vessel_interlock} passes in two of three repetitions, while \tool{mixing_tank_fill_heat} is usually a host-test failure and \tool{filter_tank_sequence} never passes.

Below 27B, the weaker local models mostly fail after producing candidate implementations rather than by failing to engage at all. For \tool{qwen3.5-35B-A3B}, \tool{qwen3.5-9B}, and \tool{qwen3.5-4B}, host-test failure is the typical outcome on \tool{thermal_chamber_hysteresis}, \tool{pressure_vessel_interlock}, and \tool{filter_tank_sequence}, while \tool{mixing_tank_fill_heat} is the task where integration-style failures become common. The smallest model, \tool{qwen3.5-2B}, sits in a different regime altogether: it never passes and its modal outcome is no submission on every task.

Taken together, the two panels suggest that the main separation below \tool{qwen3.5-27B} is not just lower search effectiveness. The harder tasks expose a more basic weakness in producing code that survives local verification, and at 2B that weakness appears one step earlier as failure to produce viable submissions at all.

\subsection{Effect of Feedback Visibility}

\begin{figure*}[t]
\centering
\includegraphics[width=\textwidth]{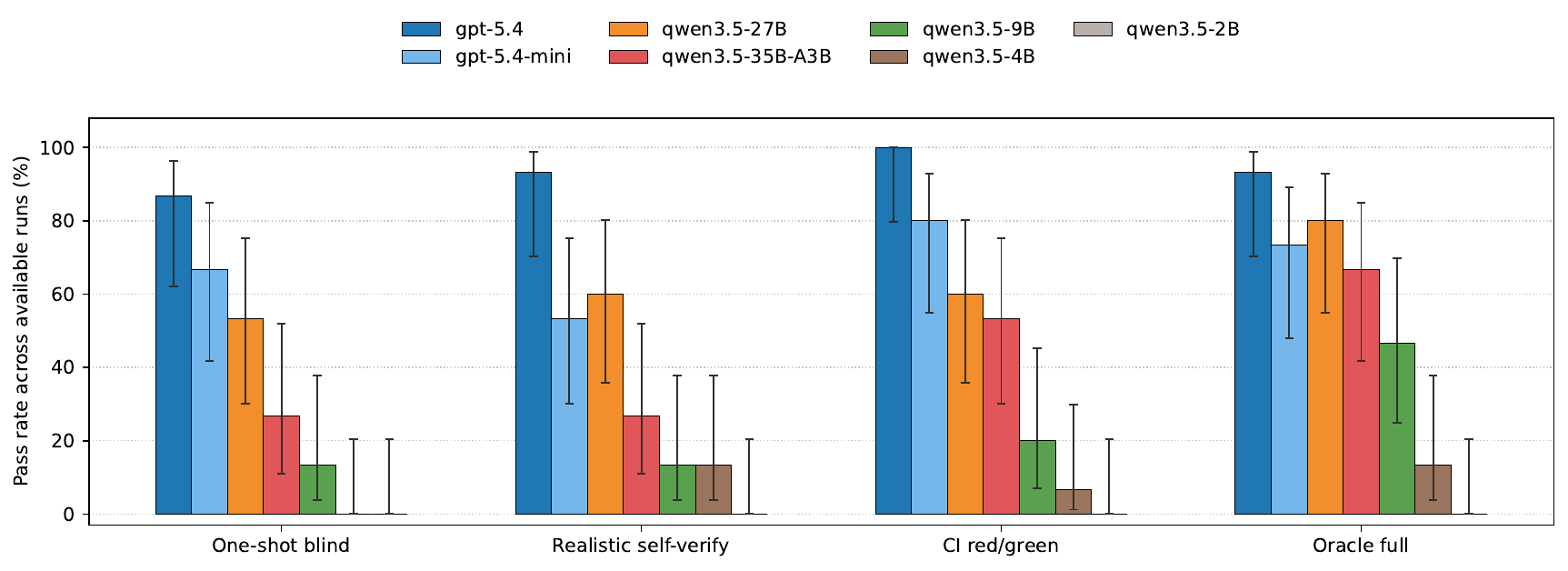}
\caption{Pass-rate sensitivity to feedback visibility across the evaluated model set, pooled over three repetitions. Each bar is a percentage over the 15 runs available for that model and scenario, and error bars show Wilson 95\% confidence intervals.}
\label{fig:scenario-sensitivity}
\end{figure*}

The full four-scenario matrix is useful because it reveals whether a model's limitation is direct implementation quality, self-verification quality, or dependence on external search signal. Figure~\ref{fig:scenario-sensitivity} summarizes pass rates by scenario for the evaluated model set.

\begin{table*}[t]
\caption{Exact pooled pass counts for the full four-scenario matrix over three repetitions. Each cell reports the number of passes out of 15 runs for that model and scenario.}
\label{tab:scenario-matrix-exact}
\footnotesize
\begingroup
\renewcommand{\arraystretch}{1.0}
\begin{tabular*}{\textwidth}{@{}P{0.18\textwidth}cccc@{}}
\toprule
Model & \tool{oneshot_blind} & \tool{realistic_self_verify} & \tool{ci_red_green} & \tool{oracle_full} \\
\midrule
\tool{gpt-5.4} & 13/15 & 14/15 & 15/15 & 14/15 \\
\addlinespace[3.5pt]
\tool{gpt-5.4-mini} & 10/15 & 8/15 & 12/15 & 11/15 \\
\addlinespace[3.5pt]
\tool{qwen3.5-27B} & 8/15 & 9/15 & 9/15 & 12/15 \\
\addlinespace[3.5pt]
\tool{qwen3.5-35B-A3B} & 4/15 & 4/15 & 8/15 & 10/15 \\
\addlinespace[3.5pt]
\tool{qwen3.5-9B} & 2/15 & 2/15 & 3/15 & 7/15 \\
\addlinespace[3.5pt]
\tool{qwen3.5-4B} & 0/15 & 2/15 & 1/15 & 2/15 \\
\addlinespace[3.5pt]
\tool{qwen3.5-2B} & 0/15 & 0/15 & 0/15 & 0/15 \\
\bottomrule
\end{tabular*}
\endgroup
\end{table*}

Also, the exact pooled counts behind Figure~\ref{fig:scenario-sensitivity} are shown in Table~\ref{tab:scenario-matrix-exact}, which lists the full four-scenario matrix directly as \textit{x}/15 values.

From these results, three patterns stand out: First, \tool{gpt-5.4} shows high but not perfect stability across feedback regimes: it reaches 13/15 (86.7\%) in \tool{oneshot_blind}, 14/15 (93.3\%) in \tool{realistic_self_verify}, 15/15 (100\%) in \tool{ci_red_green}, and 14/15 (93.3\%) in \tool{oracle_full}. The important point is therefore relative stability rather than saturation: it is the strongest model in every scenario. 

Second, \tool{qwen3.5-27B} is unusually strong in the benchmark's primary setting, not only in the rich-feedback settings. It reaches 9/15 (60.0\%) in \tool{realistic_self_verify} and 12/15 (80.0\%) in \tool{oracle_full}, versus 8/15 (53.3\%) in one-shot and 9/15 (60.0\%) in CI-style red/green search. That is useful because it shows that the model is not simply depending on external hidden guidance to recover performance.

Third, the weaker local models are much more feedback-sensitive. The clearest example is the 35B-A3B MoE model: it reaches only 4/15 in both one-shot and realistic self-verification, rises to 8/15 under CI-style red/green feedback, and reaches 10/15 under oracle feedback. This suggests that the model can often repair behavior once hidden feedback is made rich enough, but it is substantially weaker at constructing its own local verification loop. The 9B and 4B models show the same qualitative trend, although at a lower overall capability level. The 2B model does not respond meaningfully even to richer feedback: it remains flat at zero across all 60 runs.

The hosted smaller frontier model behaves differently again. \tool{gpt-5.4-mini} peaks in the CI-style condition at 12/15 rather than in oracle or realistic self-verification. That is exactly the kind of scenario-specific behavior the benchmark is supposed to reveal. The scenario axis is therefore not just an ``easier versus harder'' dial. It is a diagnostic lens on how models use, or fail to use, different forms of feedback.

\subsection{Local Models on Shared Hardware}

The local-only analysis is where wall-clock becomes meaningful. All local Qwen runs in this study use the same \tool{pi} harness, the same task packets, the same QEMU-based benchmark environment, the same \tool{llama.cpp} serving stack, and the same NVIDIA RTX 4090. We also deliberately kept the study within one GGUF 4-bit deployment regime, rather than mixing quantized checkpoints with higher-precision variants. Even though a 9B model could fit at higher precision on this hardware, we kept it quantized so that model size and architecture remain the primary varying factors in the comparison.

\begin{table*}[t]
\caption{Pooled local-only Qwen comparison on shared hardware across three repetitions. Pass denominators are 60 for every model. The pass column reports pooled counts, pass rate, and Wilson 95\% confidence interval. The \textit{n submit} column reports how many runs reached at least one submission; average first-submit time is conditioned on those runs.}
\label{tab:local-qwen-overall}
\small
\begingroup
\renewcommand{\arraystretch}{1.0}
\begin{tabular*}{\textwidth}{@{}P{0.17\textwidth}ccccccc@{}}
\toprule
Model & Form & Passes & \makecell[c]{Avg\\wall-clock} & \makecell[c]{$n$\\submit} & \makecell[c]{Avg 1st\\submit} & \makecell[c]{Avg total\\tokens} & \makecell[c]{Avg\\tool calls} \\
\midrule
\tool{qwen3.5-27B} & Dense & \makecell[c]{38/60 (63.3\%)\\[-1pt]{\scriptsize[50.7, 74.4]}} & 26.2 min & 60/60 & 18.0 min & 4.08M & 77.0 \\
\addlinespace[4.5pt]
\tool{qwen3.5-35B-A3B} & MoE & \makecell[c]{26/60 (43.3\%)\\[-1pt]{\scriptsize[31.6, 55.9]}} & 16.9 min & 60/60 & 11.9 min & 4.19M & 80.1 \\
\addlinespace[4.5pt]
\tool{qwen3.5-9B} & Dense & \makecell[c]{14/60 (23.3\%)\\[-1pt]{\scriptsize[14.4, 35.4]}} & 25.3 min & 56/60 & 17.6 min & 6.09M & 103.5 \\
\addlinespace[4.5pt]
\tool{qwen3.5-4B} & Dense & \makecell[c]{5/60 (8.3\%)\\[-1pt]{\scriptsize[3.6, 18.1]}} & 21.4 min & 50/60 & 9.4 min & 11.55M & 179.2 \\
\addlinespace[4.5pt]
\tool{qwen3.5-2B} & Dense & \makecell[c]{0/60 (0.0\%)\\[-1pt]{\scriptsize[0.0, 6.0]}} & 29.7 min & 8/60 & 1.3 min & 54.95M & 716.3 \\
\bottomrule
\end{tabular*}
\endgroup
\end{table*}

\begin{figure*}[t]
\centering
\includegraphics[width=\textwidth]{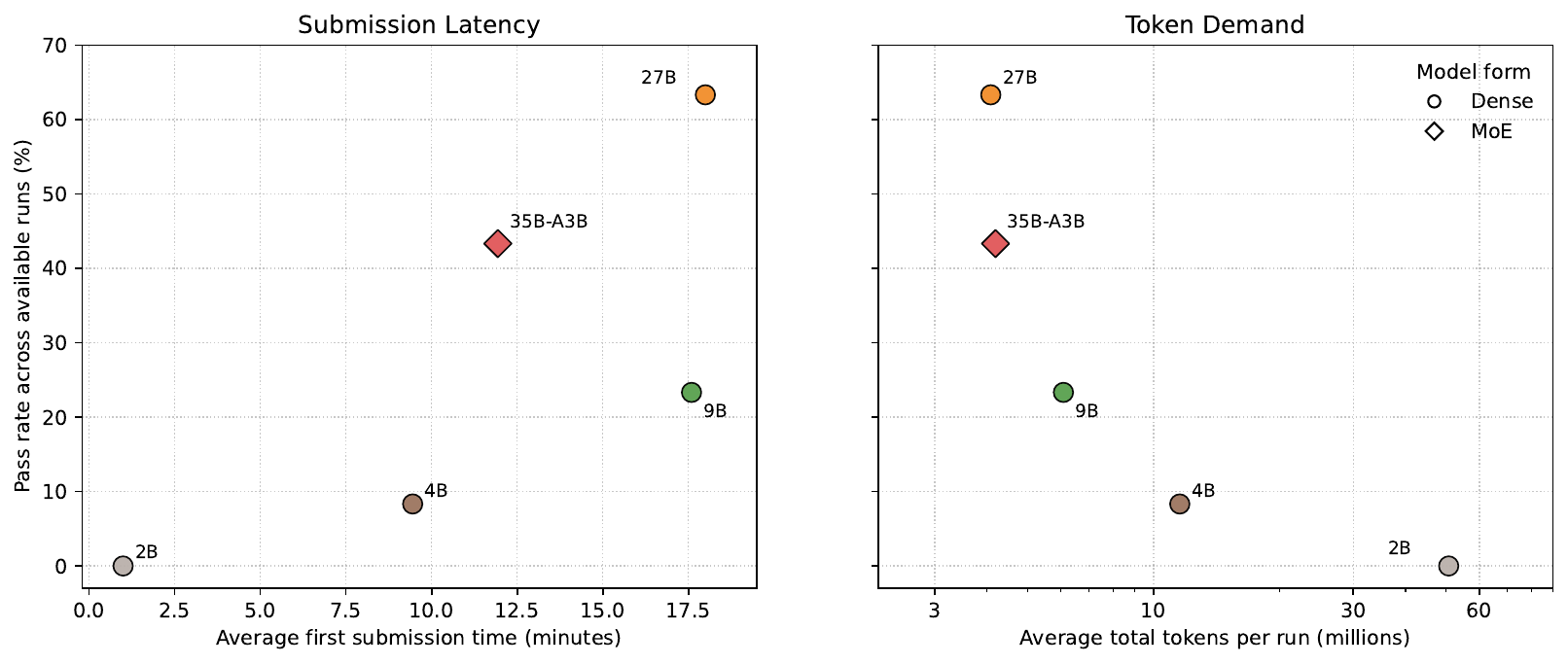}
\caption{Shared-hardware local-model tradeoffs shown as two side-by-side views. Both panels use pass rate rather than raw pass count so that accuracy and resource demand share a common scale. The left panel relates pass rate to average first-submission latency; the right panel relates pass rate to average total token demand.}
\label{fig:local-qwen-tradeoff}
\end{figure*}

Table~\ref{tab:local-qwen-overall} shows that the best local operating point is the 27B dense model. It is the most accurate local model at 38/60 (63.3\%), and it is slightly more interaction-efficient than the 35B-A3B MoE model. The 35B-A3B MoE model is the main counterexample. It reaches 26/60 (43.3\%), submits earlier on the same hardware, but is less accurate and slightly more token-hungry on average. In this benchmark, the 35B-A3B model therefore looks like a throughput advantage rather than a capability advantage. A plausible explanation is that, despite the 35B model being larger, it is a Mixture-of-Experts (MoE) model \cite{shazeer2017outrageously} in which only about 3B parameters are simultaneously active during inference.

The 2B model sharpens the lower boundary of the family even more. Despite being the smallest checkpoint, it is not the cheapest operating point in closed-loop use: averaged over its 60 pooled runs it consumes 54.95M tokens, 716.3 tool calls, and 29.7 minutes per run while producing zero passes. Its reported 1.3-minute time to first submission should not be read as an efficiency signal, because only eight of its 60 runs recorded a first submission in the final curated bundle.

Figure~\ref{fig:local-qwen-tradeoff} makes the same ranking visible visually in a way that separates the two resource axes. The left panel shows that the faster first-submission models are not necessarily the better ones: the 35B-A3B model submits earlier than 27B, but converts less of its available runs into success, while 2B submits almost immediately in the few runs that submit at all yet never converts that into success. The right panel isolates the token story even more clearly. As model scale decreases below the practical operating range, token demand rises sharply rather than falling: 4B is already far to the right of 27B and 35B-A3B, and 2B remains an extreme outlier. In other words, the failure mode at the lower end of this family is uncontrolled search rather than simple throughput loss.

\rev{The smaller dense models still degrade sharply. The 9B model remains meaningful at 14/60 (23.3\%), but it already requires far heavier search than the 27B model, averaging 103.5 tool calls per run. The 4B model is close to the lower useful boundary of the current benchmark: it reaches only 5/60 runs (8.3\%), yet averages 11.55M tokens and 179.2 tool calls. Its shorter average time to first submission is not evidence of efficiency; it is largely evidence of premature submission followed by repeated failure-driven search. The 2B model lies beyond that boundary: across 60 runs, it reaches zero passes, never solves any task in any scenario, and is dominated by long no-submission loops rather than productive repair.}

The task-wise pattern points in the same direction. \tool{filter_tank_sequence} remains the clearest local separator in the suite: across the twelve local runs per model-task pair, it is solved only once by \tool{qwen3.5-27B} and never by the smaller local models. \tool{tank_fill_drain}, by contrast, remains the easiest local task and is solved at least once by every local model down to 4B. The most deployment-relevant comparison remains between the 27B dense model and the 35B-A3B MoE model. The 35B model matches or exceeds the 27B model only on \tool{tank_fill_drain}; it is weaker on \tool{thermal_chamber_hysteresis}, \tool{pressure_vessel_interlock}, \tool{mixing_tank_fill_heat}, and \tool{filter_tank_sequence}. This is a strong indication that sparse-parameter throughput does not automatically translate into stronger autonomous embedded-agent behavior.

The lower end of the family points in the same direction even more sharply. The 2B model is 0/12 on every task. Across all 60 pooled cells, it produces no passes and records a first submission in only eight runs. We therefore treat 2B as the lower boundary of the useful local range rather than as a serious contender in the ranking, and we do not continue further downward to 0.8B in the current study.

\section{Discussion}
\label{sec:discussion}

The evaluation results support three main interpretive claims. First, the primary \tool{realistic_self_verify} condition is the most informative lens for judging autonomous embedded-agent capability. Second, the feedback-visibility axis is diagnostically useful because it separates self-verification from externally guided repair. Third, the shared-hardware local study changes the deployment story by showing how model scale interacts with latency, token demand, and search behavior.

\rev{The most important model-separation pattern in the paper is the ordering observed under \tool{realistic_self_verify}.} That condition is the closest match to the intended benchmark use case, because the agent must rely on the visible build, test, and runtime surface rather than on hidden diagnostics. Under that condition, \tool{gpt-5.4} leads at 14/15 (93.3\%), \tool{qwen3.5-27B} posts the highest pooled count among non-frontier models at 9/15 (60.0\%), \tool{gpt-5.4-mini} reaches 8/15 (53.3\%), the 35B/9B/4B local variants reach 4/15 (26.7\%), 2/15 (13.3\%), and 2/15 (13.3\%) respectively, and \tool{qwen3.5-2B} remains 0/15 while usually failing to submit at all.

\rev{This ordering matters as a descriptive result because it reveals a clear capability gradient under the benchmark's primary condition.} It distinguishes models that can genuinely construct and trust local evidence from models that mainly recover when external search signal is made richer. It also highlights a practically important point: local open-weight agents are already nontrivial contenders in the benchmark's primary setting. The best local model is not at parity with the frontier model, but it is clearly beyond a toy baseline. That is exactly the kind of deployment-relevant gradient the benchmark was designed to expose.

The scenario axis is best interpreted as a diagnostic tool rather than as a single monotone leaderboard. \tool{gpt-5.4} shows strong but not perfect stability across feedback visibility and is strongest in every scenario. Below that level, models react very differently to additional hidden feedback. \tool{gpt-5.4-mini} improves most under CI-style red/green feedback. \tool{qwen3.5-27B} is strongest in \tool{oracle_full} and stays competitive in \tool{realistic_self_verify}, which suggests that its relative strength is not merely hidden-feedback dependence. By contrast, the 35B-A3B MoE model improves sharply as hidden feedback becomes richer, which indicates that it can often repair once the evaluator points it in the right direction, but struggles to build a strong self-verification loop on its own.

This is important for benchmark design. If the paper reported only one-shot generation, it would miss whether a model can actually engage in an engineering loop. If it reported only oracle-style repair, it would overstate autonomous capability. Using all four scenarios makes it possible to separate direct implementation skill, self-verification skill, and feedback dependence.

The local-only analysis adds a second practical lesson. Once hardware and serving are controlled, wall-clock becomes meaningful, and the resulting trade-offs are not the same as the cross-backend headline. On a single RTX 4090 with \tool{llama.cpp}, the best local operating point is the 27B dense model at 38/60 runs. The 35B-A3B MoE model is faster to first submit, but it is not better on this benchmark. Its behavior is consistent with a throughput advantage rather than a capability advantage.

The smaller dense models are also informative. The 9B model is a meaningful, if clearly weaker, embedded agent. The 4B model appears close to the lower useful boundary of the benchmark, because it spends very large interaction budgets for relatively few successes. The 2B result places that lower boundary more firmly still: despite being the smallest checkpoint, it is slower and far more token-hungry in practice because it falls into long no-submission loops. If a model cannot produce passes, or even regular submissions, across the suite, then it is below the current useful operating range of the benchmark. This same-hardware analysis is important because it answers a practical question that hosted-versus-local comparisons cannot: which local model family member is actually worth deploying under a fixed on-premises budget.

Our choice to keep the local study within a common 4-bit GGUF deployment regime is also part of that interpretation. We did not want a precision change to masquerade as a model-size effect. A separate higher-precision ablation, especially around the 9B scale, would be useful in future work, but it would answer a different question from the one addressed here.

Several benchmark-design lessons follow directly from the pooled results. First, \tool{filter_tank_sequence} is the main separator in the current suite and remains difficult even when strong models pass it, so it is doing the right kind of work. Second, \tool{tank_fill_drain} functions well as a baseline anchor because even weak models can sometimes solve it.

The benchmark still has important scope limits that matter for interpretation. It studies deterministic simulated plants on a single ESP32/QEMU reference stack with UART-virtualized I/O, which is a deliberate choice for reproducibility and comparability rather than an attempt to cover the full complexity of embedded deployment. Likewise, the current suite is intentionally small and control-centric: the five tasks are enough to separate models on thresholding, hysteresis, sequencing, and interlocks, but they do not yet span the broader range of embedded workloads such as communication stacks, storage, power management, interrupt-heavy firmware, or longer-horizon mission logic. The evaluation also fixes one agent/tool surface and one common local deployment regime, so the reported rankings should be read as benchmarked results in this setting rather than as architecture-independent limits of all coding agents. However, these scope choices are also what make the study useful: they turn a noisy and difficult problem into a reproducible one, and they already produce clear capability gradients, informative failure modes, and actionable deployment trade-offs. On that basis, we view the present paper not as a final word on embedded-agent capability, but as a strong benchmark foundation on which broader and more realistic evaluations can now be built.

\section{Conclusions and Future Work}
\label{sec:conclusion}

This paper presents a closed-loop benchmark methodology for evaluating coding agents on embedded firmware tasks, where success is defined at the level of observed device behavior rather than source-code plausibility alone. Starting from prose engineering handoffs, the benchmark asks whether an agent can iteratively implement, verify, diagnose, and repair firmware under a realistic tool-mediated workflow. In that sense, the main contribution is not only a five-task suite, but an evaluation framing: prose-first, behavior-level, and explicitly centered on self-verification in a target execution environment.

The empirical study shows that this benchmark is already meaningful along several dimensions. It is executable at scale, reproducible under a deterministic simulation stack, and sufficiently discriminative to separate models both by final pass rate and by how they use feedback during iterative search. Across 420 runs, the primary \tool{realistic_self_verify} condition emerges as the most informative lens on autonomous embedded-agent capability, because it requires the agent to construct and trust its own local evidence rather than relying on hidden evaluator guidance. Under that condition, \rev{\tool{gpt-5.4} has the highest observed pass rate} at 14/15 (93.3\%), while \rev{\tool{qwen3.5-27B} is the strongest observed local model} at 9/15 (60.0\%). This result is notable in practical terms: the 27B local model remains genuinely competitive in the benchmark's primary setting despite operating in a far smaller and more resource-constrained deployment regime than the hosted frontier system.

A second conclusion is that feedback visibility is not just a convenience variable but a diagnostic one. The four-scenario design shows that models differ not only in direct implementation ability, but also in how well they construct self-verification loops and how strongly they depend on richer hidden feedback to repair failures. Some models remain comparatively stable across scenarios, whereas others improve sharply only when the evaluator exposes richer search signal. This makes the feedback axis scientifically useful in its own right: it helps separate first-pass synthesis, self-verification quality, and externally guided repair instead of collapsing them into a single aggregate score.

A third conclusion is practical. Under a common local deployment regime on shared hardware, the best local operating point in this study is the 27B dense Qwen model. The comparison with the \tool{qwen3.5-35B-A3B} MoE model is especially informative: the MoE model reaches first submission faster, but the dense 27B model is clearly stronger on the benchmark overall. In this setting, the contrast is best interpreted as a trade-off between throughput and effective capability. Faster submission does not automatically translate into stronger autonomous embedded-agent performance, and sparse activation does not substitute for the ability to sustain reliable closed-loop search.

The lower end of the local family reinforces the same lesson. Smaller local models degrade not only in pass rate but also in search efficiency, with the 2B checkpoint showing that lower nominal model size does not necessarily produce a cheaper or more usable agent when closed-loop interaction is taken seriously. In this benchmark, the key deployment boundary is therefore not simply whether a model can emit plausible code, but whether it can reliably converge through repeated verification and revision without falling into unproductive search.

Taken together, these results suggest that closed-loop embedded-agent evaluation should be treated as a distinct benchmarking problem rather than as a small variation on static code generation. The main scientific gap below frontier performance is not only implementation quality, but robust autonomous self-verification: the ability to generate local evidence, interpret it correctly, and stop only when the controller is actually ready. That is precisely the capability this benchmark is designed to expose.

Regarding future work, the most important next step is to carry the same methodology from deterministic simulation into hardware-in-the-loop and then carefully controlled real-device studies. This is the main long-term target of the benchmark, even though it is substantially harder experimentally, because only real hardware can fully surface effects such as sensor jitter, communication latency, watchdog behavior, startup transients, peripheral interaction, and deployment packaging. The role of the present simulation-based study is therefore methodological staging: it establishes a reproducible baseline from which hardware-facing evaluation can proceed in a controlled way.

At the same time, research on local models is especially promising for embedded-agent workflows because it offers direct control over both data and compute, which is highly relevant for privacy-sensitive, on-premises, and industrial settings. Future work should therefore continue to study local-model operating points, precision regimes, and agent/tool surfaces, alongside broader task extensions that include noisier and partially missing sensors, delayed or drifting signals, actuator saturation, restart-and-recovery behavior, multi-rate control loops, and richer fault-management scenarios. Together, these directions would turn the present benchmark from a single reproducible suite into a broader evaluation program for trustworthy embedded coding agents.

\section*{Acknowledgements}

This work was supported by the AgroVAL (TSI-100122-2024-10) project funded by the Recovery, Transformation and Resilience Plan from the European Union Next Generation through the Ministry for Digital Transformation and the Civil Service, the KOSMOS-UA project (PID2024-155363OB-C43), funded by the Spanish Ministry of Science and Innovation, BALIDA-AA project (CIPROM/2024/13), funded by Conselleria de Educación, Cultura, Universidades y Empleo (Generalitat Valenciana), the IAEAV project (INREIA/2024/176) funded by the Conselleria de Innovación, Industria, Comercio y Turismo (Generalitat Valenciana), the ENIA Chair of Artificial Intelligence from the University of Alicante (TSI-100927-2023-6) funded by the Recovery, Transformation and Resilience Plan from the European Union Next Generation through the Ministry for Digital Transformation and the Civil Service, and the Grant RED2022-134656-T funded by MCIN/AEI/10.13039/501100011033.

Sergio García-Carrasco holds a predoctoral grant of the ACIF program (CIACIF/2023/019) funded by Generalitat Valenciana and the European Union's ESF+.

\section*{CRediT Authorship Contribution Statement}

Jorge García-Carrasco: Conceptualization, methodology, software, validation, formal analysis, investigation, writing --- original draft, writing --- review and editing. Sergio García-Carrasco: Conceptualization, methodology, software, validation, writing --- review and editing. Alejandro Maté: Supervision, project administration, writing --- review and editing. Juan Trujillo: Supervision, project administration, writing --- review and editing.

\section*{Declaration of Competing Interest}

The authors declare that they have no known competing financial interests or personal relationships that could have appeared to influence the work reported in this paper.

\section*{Data Availability}

The benchmark repository, task packets, runner scripts, configuration files, and curated evaluation artifacts are publicly available at \url{https://github.com/jgcarrasco/closed_loop_evaluation_agents_embedded}. The paper-facing artifact bundle is organized under \tool{artifacts/evaluations/paper_closed_loop_eval_420/} and includes per-run prompts, session transcripts, runner logs, and summary files for all 420 official runs.

\bibliographystyle{unsrtnat}
\bibliography{references}

\end{document}